 \documentclass[final,5p,times,twocolumn]{elsarticle}
  \usepackage{graphicx}

\usepackage{amssymb}
\usepackage{textcomp}

\biboptions{square,numbers,sort&compress}

\journal{Computational Materials Science}
\begin{document}

\begin{frontmatter}



\title{Force and magnetic field sensor based on measurement of tunneling conductance between ends of coaxial carbon nanotubes}


\author[ISRAN]{Andrey M. Popov}
\ead{popov-isan@mail.ru}
\author[NanoBioSpectroscopy]{Irina V. Lebedeva}
\author[KintechLab,KurchatovInstitute]{Andrey A. Knizhnik}
\author[ISRAN,MIPT]{Yurii~E.~Lozovik}
\author[BSU]{Nikolai A. Poklonski\corref{cor}}
\ead{poklonski@bsu.by}
\cortext[cor]{Corresponding author}
\author[BSU]{Andrei I. Siahlo}
\author[BSU]{Sergey A. Vyrko}
\author[BSU]{Sergey V. Ratkevich}

\address[ISRAN]{Institute of Spectroscopy Russian Academy of Science, Fizicheskaya Str. 5, Troitsk, Moscow 142190, Russia}
\address[NanoBioSpectroscopy]{Nano-Bio Spectroscopy Group and ETSF Scientific Development Centre, Departamento de Fisica de Materiales, Universidad del Pais Vasco UPV/EHU, Avenida de Tolosa 72, E-20018 San Sebastian, Spain}
\address[KintechLab]{Kintech Lab Ltd., Kurchatov Sq. 1, Moscow 123182, Russia}
\address[KurchatovInstitute]{National Research Centre ``Kurchatov Institute'', Kurchatov Sq. 1, Moscow 123182, Russia}
\address[MIPT]{Moscow Institute of Physics and Technology, Institutskii pereulok 9, Dolgoprudny, Moscow Region 141700, Russia}
\address[BSU]{Physics Department, Belarusian State University, pr. Nezavisimosti 4, Minsk 220030, Belarus}


\begin{abstract}
The interaction and tunneling conductance between oppositely located ends of coaxial carbon nanotubes are studied by the example of two (11,11) nanotubes with open ends terminated by hydrogen atoms. The Green function formalism is applied to determine the tunneling current through the nanotube ends as a function of the distance between the ends, relative orientation of the nanotubes and voltage applied. The energy favorable configuration of the coaxial nanotubes is obtained by the analysis of their interaction energy at different distances between the nanotube ends and angles of their relative rotation. Using these calculations, a general scheme of the force sensor based on the interaction between ends of coaxial nanotubes is proposed and the relation between the tunneling conductance and measured force is established for the considered nanotubes. The operational characteristics of this device as a magnetic field sensor based on measurements of the magnetic force acting on the coaxial nanotubes filled with magnetic endofullerenes are estimated.
\end{abstract}

\begin{keyword}
Carbon nanotubes \sep
Tunneling conductance \sep
Magnetic endofullerenes \sep
Magnetic force \sep
Force sensor
\end{keyword}

\end{frontmatter}



\section{Introduction}
\label{Introduction}

Unique elastic properties and metallic conductivity of carbon nanotubes allow using the nanotubes as parts of
nanoelectromechanical systems (NEMS) (see Refs. \cite{Dong07, LozovikPopov07PhysUsp, Bichoutskaia08} for a
review). The discovery of relative motion of the walls \cite{Cumings00} in multiwalled carbon nanotubes
(MWNTs) was immediately followed by the idea that NEMS can be based on such a motion \cite{Forro00}. In the
last decade, a number of NEMS based on the relative motion of nanotube walls have been implemented
experimentally. Among these devices, there are nanomotors in which walls of a MWNT play roles of the shaft
and bush driven by an electric field \cite{Fennimore03, Bourlon04, Subramanian07} or a thermal gradient
\cite{Barreiro08} and memory cells operating on relative sliding of the walls along the nanotube axis
\cite{Deshpande06, Subramanian10}. A wide set of such NEMS have been also proposed and studied theoretically,
including a gigahertz oscillator \cite{ZhengJiang02, ZhengLiuJiang02}, an accelerometer \cite{WangJiang08,
KangLeeKimChoi09}, a nanothermometer \cite{Bichoutskaia07}, an ultrahigh frequency resonator based on the
relative vibrations of the nanotube walls \cite{Bichoutskaia09}, a bolt/nut pair \cite{Barreiro08, Saito01,
LozovikMinoginPopovPhysLett03, LozovikMinoginPopovJETPLett03}, a nanoactuator in which a force directed along
the nanotube axis leads to rotational motion of the walls \cite{PopovBichoutskaiaLozovikKulish07} and a
scanning rotational microscope \cite{PopovLebedevaKnizhnik12}.

\begin{figure*}[!t]
\center{\includegraphics
{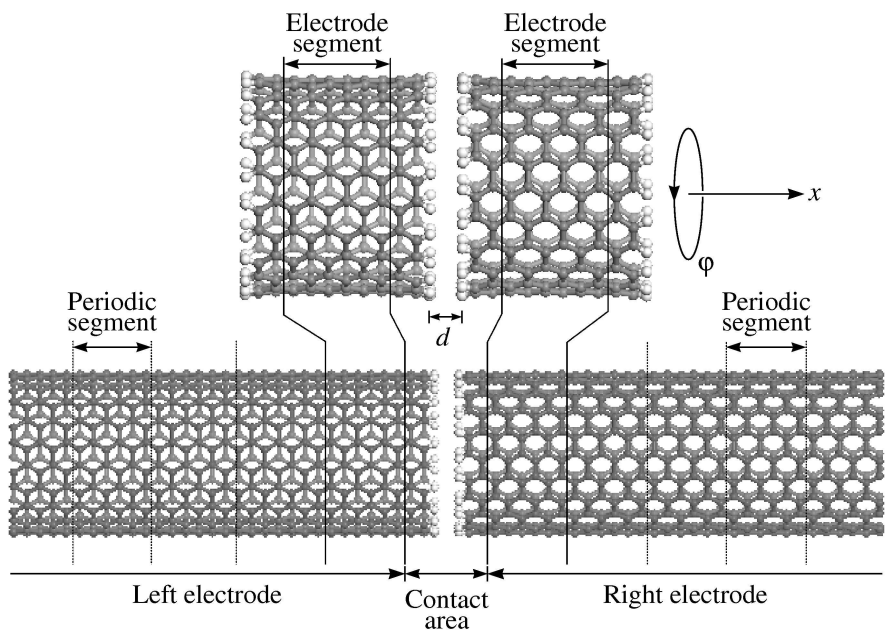}}
\caption{Atomic model used for calculations of electron transport and van der Waals interaction between coaxial (11,11) carbon nanotubes with the edges terminated by hydrogen atoms. (Below) The system with semi-infinite (11,11) carbon nanotube electrodes for which electron transport calculations are performed. (Above) The finite (11,11) carbon nanotubes used for calculations of the van der Waals interaction and electronic structure of the contact area of the nanotubes and its coupling to the electrodes. The vertical solid lines denote the contact area and segments of the electrodes within the finite model of the nanotube contact. The vertical dotted lines denote periodic segments of the semi-infinite electrodes that are used in the calculations. Carbon and hydrogen atoms are colored in gray and white, respectively. $\varphi$ is the angle of relative rotation of the nanotubes around the common axis.}
\label{figures:Fig01AtomicModel}
\end{figure*}

The characteristics of the memory cells mentioned above \cite{Deshpande06, Subramanian10}, such as the
conductance in the ``ON'' state, operational frequency, and possibility to use the cells in volatile or
non-volatile memory, are determined by interaction and conductance between oppositely located ends of coaxial
carbon nanotubes. Thus, theoretical modeling of the interaction and conductance between nanotube ends holds
the key to success of these applications. In the present paper, atomistic calculations of the interaction
energy between ends of coaxial carbon nanotubes using semi-empirical potentials are accompanied by
calculations of the tunneling conductance in the framework of the Green function formalism. The results of
these calculations can be used for simulation and comprehensive analysis of operational characteristics of
nanotube-based NEMS and nanoelectronic devices. A set of force and mass sensors based on measurements of
frequency of string-like vibration of carbon nanotubes have been implemented \cite{Sazonova04, Peng06,
Lassagne08}. The sensor based on such measurements was proposed to be used to determine magnetic moment of a
nanoobject attached to a nanotube \cite{Lassagne11}. The change of nanotube conductance at nanotube torsion
\cite{Cohen-Karni09} or bending \cite{Stampfer06} and dependence of the tunneling conductance between
nanoobjects at their relative displacement at subangstrom scale \cite{Bichoutskaia07, Lam09, Poklonski13} can
be used for NEMS elaboration. Particularly a nanothermometer based on measurements of the tunneling
conductance between walls of a double-walled carbon nanotube \cite{Bichoutskaia07} and a nanodynamometer
based on measurements of the tunneling conductance between adjacent graphene layers \cite{Lam09, Poklonski13}
have been proposed.
In~\cite{Kibalchenko11} the shielding of external axial constant magnetic field by induced currents in zigzag $(n,0)$ nanotubes is examined theoretically. Enhancement (paramagnetic response of nanotube for $n$ multiply of 3) or reduction (diamagnetic response of nanotube for $n$ aliquant of 3) of external magnetic field is manifested in induced chemical shifts of NMR signal from molecules encapsulated in nanotube. In principle, the chemical shift of NMR signal allows to estimate the value of external magnetic field.
Here we develop a general scheme and operational principles of the force sensor based both on relative motion
of carbon nanotubes walls and measurements of the tunneling conductance between their ends.

We also suggest that filling of coaxial nanotubes in the proposed force sensor with magnetic endofullerenes can be used for measurements of magnetic fields. A variety of endofullerenes and nanotubes filled with fullerenes (nanotube peapods), including nanotube peapods filled with magnetic endofullerenes \cite{Hirahara00, Shiozawa06, SunInoueShimada04, KitauraOkimoto07}, can be obtained in macroscopic amounts \cite{KitauraShinohara07}. It has been also shown that the magnetic moment is greater for magnetic endofullerenes inside carbon nanotubes than for the same magnetic endofullerenes when they are isolated \cite{KitauraOkimoto07}. Recently operational characteristics of the magnetic nanorelay based on bending of the nanotubes filled with the magnetic endofullerenes have been calculated \cite{PoklonskiKislyakov10}. Here we consider the possibility to determine a magnetic field through measurements of the magnetic force between two coaxial nanotubes filled with the magnetic endofullerenes using the proposed force sensor. Up to now the largest magnetic moment of 21 Bohr magnetons has been observed for (Ho$_3$N)@C$_{80}$ \cite{Wolf05}. The operational characteristics of the magnetic field sensor based on the coaxial (11,11) nanotubes filled with the (Ho$_3$N)@C$_{80}$ magnetic endofullerenes with the largest observed magnetic moment
are calculated.

The paper is organized in the following way. In Section \ref{VanDerWaalsSection}, we determine the optimal configuration of the coaxial nanotubes by consideration of the van der Waals interaction between their ends. In Section \ref{TunnelingConductance}, the results of calculations of electron tunneling current between the ends of coaxial carbon nanotubes as a function of the distance between them and the voltage applied are given. Section \ref{ForceSensorSection} is devoted to the general scheme, operational principles and characteristics of the considered system as a force sensor. In Section \ref{MagneticForceSection}, we discuss the possibility to use this sensor for detection of magnetic fields through measurements of magnetic forces. Our conclusions are summarized in Section \ref{ConclusionsSection}.

\section{Van der Waals interaction calculations}
\label{VanDerWaalsSection}

To start consideration of tunneling between coaxial carbon nanotubes we have first determined their energy favorable configuration. Calculations of interaction energy between two coaxial carbon nanotubes (Fig.~\ref{figures:Fig01AtomicModel}) have been performed using in-house MD-kMC (Molecular Dynamics - kinetic Monte Carlo) code \cite{kintechlab}. Two (11,11) single-walled carbon nanotubes of 12.9~\AA{} length with both edges passivated with hydrogen atoms are separately geometrically optimized using the Brenner potential \cite{Brenner02} (the binding energy of the nanotubes and its dependence on relative orientation of the nanotube change by no more than 1\% and 2\%, respectively, upon increasing the nanotube length twice). Then the walls are considered as rigid, brought together in the coaxial configuration, shifted along the axis and rotated with respect to each other. The interaction between carbon nanotubes is described using the Lennard-Jones 12-6 potential
\begin{equation}\label{eq101}
   U(r) = 4\varepsilon\left[\left(\frac{\sigma}{r}\right)^{12} - \left(\frac{\sigma}{r}\right)^{6}\right],
\end{equation}
where  $\sigma_\mathrm{CC} = 3.40$~\AA{}, $\sigma_\mathrm{HH} = 2.60$~\AA{}, and $\sigma_\mathrm{CH} = 3.00$~\AA{}, $\varepsilon_\mathrm{CC} = 3.73$~meV, $\varepsilon_\mathrm{HH} = 0.650$~meV and $\varepsilon_\mathrm{CH} = 1.56$~meV for carbon-carbon, hydrogen-hydrogen and carbon-hydrogen interactions, respectively \cite{Cornell95}.

\begin{figure}[!t]
\center{\includegraphics
{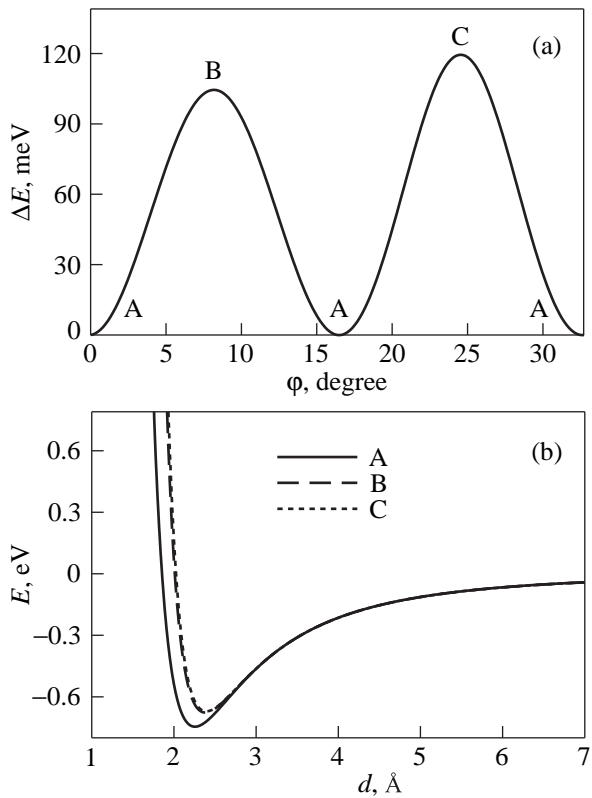}}
\caption{(a) Calculated changes $\Delta E$ in the interaction energy of two coaxial (11,11) carbon nanotubes with the edges terminated by hydrogen atoms upon their relative rotation around the common axis by the angle $\varphi$ for the distance between the nanotube ends of $d_0 = 2.26$ \AA{} corresponding to the minimum of the interaction energy as a function of the distance $d$ between the nanotube ends.
The angle $\varphi = 0^\circ$ corresponds to the configuration shown in Fig.~\ref{figures:Fig01AtomicModel}. The relative orientations corresponding to minima and maxima are denoted by A and by B and C, respectively. (b) Calculated interaction energy $E$ of the nanotubes as a function of the distance $d$ between the nanotube ends along the axis for different relative orientations of the nanotubes: A (solid line), B (dashed line), C (dotted line).}
\label{figures:Fig02CalculatedChangesE}
\end{figure}

First the global energy minimum for two rigid nanotubes has been found. This energy minimum corresponds to the distance $d$ between the nanotube ends along the nanotube axis of $d_0 = 2.26$~\AA{}
and the relative orientation of the nanotubes shown in Fig.~\ref{figures:Fig01AtomicModel}. The binding energy of the nanotubes is found to be $-0.746$~eV (Fig.~\ref{figures:Fig02CalculatedChangesE}b). The energy favorable orientation of the carbon nanotubes corresponds to the case when hydrogen atoms of one of the nanotubes are located in the middle between hydrogen atoms of the neighboring edge of the second nanotube, i.e. at the given distance between the nanotubes, the hydrogen atoms tend to be as far as possible from each other. For the distance between the nanotube ends of $d_0 = 2.26$~\AA{},
the distances between adjacent hydrogen atoms of different nanotubes lie in the range from 2.4~\AA{} to 2.7~\AA{}. There are two such minima per rotational period of the (11,11) nanotubes (Fig.~\ref{figures:Fig02CalculatedChangesE}a). We denote this relative orientation of the carbon nanotubes as A and the corresponding angles as $\varphi = 0^\circ$, $(360/22)^\circ$, $(360/11)^\circ$, etc.

The dependence of interaction energy of the carbon nanotubes on their relative orientation for the distance between the nanotube ends of $d_0 = 2.26$~\AA{}
has two types of maxima (Fig.~\ref{figures:Fig02CalculatedChangesE}a). The highest maximum C with the relative energy of 0.12 eV corresponds to the configuration in which hydrogen atoms of the neighboring edges of the nanotubes are located exactly opposite to each other. In the lower maximum B, the nanotubes are oriented in such a way that the carbon network of one of the nanotubes continues the carbon network of the second nanotube as if one long nanotube was cut into two pieces. In this maximum, adjacent hydrogen atoms are also located close to each other (at distances 2.28~\AA{}) and the relative energy of this maximum of 0.105 eV is comparable to that of maximum C. However, the optimal distances for two orientations B and C are slightly greater than the one for orientation A, 2.38~\AA{} and 2.39~\AA{}, respectively (Fig.~\ref{figures:Fig02CalculatedChangesE}b). With account of this, the barriers to relative rotation of the nanotubes via orientations B and C are found to be 0.069 eV and 0.075 eV, respectively.
These values of the barriers to relative rotation of the nanotubes are considerably larger than $k_\mathrm{B}T$ ($T$ is temperature, $k_\mathrm{B}$ is the Boltzmann constant) at liquid helium temperature, therefore the relative orientation of the carbon nanotubes should be completely frozen at this temperature. At room temperature, the relative rotation of the nanotubes becomes possible. Nevertheless, as shown below, this rotation has a negligible effect on the tunneling conductance between them.

\section{Tunneling conductance calculations}
\label{TunnelingConductance}

The tunneling conductance between the ends of the coaxial (11,11) carbon nanotubes has been studied using the non-perturbative approach based on the Green function formalism for systems with semi-infinite periodic electrodes \cite{Fonseca03}. In the case when the localized basis set is used, the system can be divided into segments with the size greater than the doubled atomic cutoff radius that only interact with the neighboring segments. Thus, the Hamiltonian of such a system has a block tridiagonal structure with segment Hamiltonians on the diagonal line and coupling matrices on the two adjacent lines. As such segments, it is convenient to choose the contact area of the nanotubes and sufficiently long periodic segments of the electrodes. Then the system is fully described by the Hamiltonians, coupling matrices and overlap matrices of periodic segments of the electrodes, the Hamiltonian of the contact area and the coupling and overlap matrices between the contact area and periodic segments of the electrodes. These matrices can be obtained by consideration of three atomic structures with a relatively small number of atoms in a simulation box (Fig.~\ref{figures:Fig01AtomicModel}): two periodic structures describing the electrodes and a finite structure comprising the contact area and one segment of each electrode. To model the periodic electrodes we take segments of the nanotubes of 7.4~\AA{} length (consisting of three elementary unit cells) so that non-adjacent segments of the electrodes do not interact. These segments are placed in 7.4~\AA{}${}\times{}$25~\AA{}${}\times{}$25~\AA{} simulation boxes with the periodic boundary conditions along the nanotube axis. The role of the finite model of the nanotube contact is played by two coaxial (11,11) carbon nanotubes of 12.9~\AA{} length with the edges terminated by hydrogen atoms. The nanotubes are placed in a 60~\AA{}${}\times{}$25~\AA{}${}\times{}$25~\AA{} simulation box. This model includes the contact area consisting of one edge elementary unit cell of each nanotube terminated with hydrogen atoms, one segment of each electrode and one more edge elementary unit cell of each nanotube to set properly boundary conditions for the periodic parts of the electrodes.

The calculations are performed in two steps. First the Hamiltonian, overlap and density matrices of the finite contact model and the periodic electrodes are obtained self-consistently in the framework of the density functional theory using OpenMX3.6 \cite{openmx-square}. The basis sets of s2p2d1 pseudo-atomic orbitals (PAOs) \cite{Ozaki03, Ozaki04} with the cutoff distance of 7 Bohr radii (C7.0-s2p2d1 and H7.0-s2p2d1) and fully relativistic pseudopotentials (CA11) are taken from the 2011 database. The exchange-correlation term is included within the local density approximation (LDA) using the Ceperley-Alder functional \cite{Ceperley80} as parameterized by Perdew and Zunger \cite{Perdew81}. The energy convergence tolerance is $2\cdot10^{-6}$~Ry (1~Ry${}= 13.606$~eV). One $\Gamma$-point and $8\times 1\times 1$ $k$-point grids are used for the calculations of the finite contact model and the periodic electrodes, respectively. The energy cutoff is 50~Ry.

At the second step, the calculated electronic structure of the finite contact model and the periodic electrodes are used to calculate the tunneling current between the semi-infinite (11,11) nanotubes using in-house Transport code based on the Green function formalism \cite{Fonseca03}. The potential is supposed to be constant at each electrode. The temperature is set to 300 K.

\begin{figure}[!t]
\center{\includegraphics
{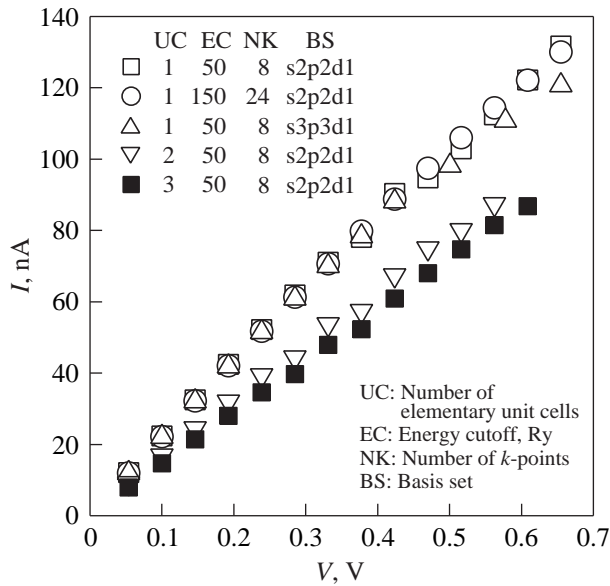}}
\caption{Calculated tunneling current $I$ between the ends of the coaxial (5,5) nanotubes as a function of voltage $V$ for relative orientation A of the nanotubes at the distance $d = 2.50$~\AA{} between the nanotube ends at temperature 300~K. In the legend the number of elementary unit cells of each nanotube in the contact area, energy cutoff, number of $k$-points along the nanotube axis for periodic electrodes, and basis set used for calculations are indicated.}
\label{figures:Fig03CalculatedTunnelingCurrent5}
\end{figure}

The convergence with respect to the energy cutoff and the number of $k$-points has been tested by the example of (5,5) nanotubes. It has been shown that the increase of the energy cutoff from 50~Ry to 150~Ry and the increase in the number of $k$-points for the periodic electrodes from $8\times1\times1$ to $24\times1\times1$ leads to small changes in the tunneling conductance to within 3\% (Fig.~\ref{figures:Fig03CalculatedTunnelingCurrent5}). The increase of the basis set from s2p2d1 to s3p3d1 results in decrease of the tunneling conductance within 2\% at voltages below 0.4~V (Fig.~\ref{figures:Fig03CalculatedTunnelingCurrent5}). The convergence on the size of the contact area is less encouraging. Inclusion of two instead of one elementary unit cells of each nanotube to the contact area leads to the decrease of the tunneling current by 25\% (Fig.~\ref{figures:Fig03CalculatedTunnelingCurrent5}). 
Inclusion of three elementary unit cells of each nanotube to the contact area results in further decrease of the tunneling current by 15\%. It should be noted that for the (11,11) nanotubes the effect of the contact area size is of the opposite sign. Inclusion of two instead of one elementary unit cell of each nanotube to the contact area leads to the increase of the tunneling current by 25\%. As we confirm below by comparison of tunneling currents for the (5,5) and (11,11) nanotubes, the results obtained using three elementary unit cells in the contact area for the (5,5) nanotubes and two elementary unit cells for the (11,11) nanotubes are sufficiently converged. Nevertheless, such calculations are too expensive to perform extensive studies for the (11,11) nanotubes. Since the relative changes in the tunneling current with increasing the contact area size are found to be the same for different distances between the nanotube ends, the results obtained below using only one elementary unit cell of each (11,11) nanotube in the contact area are scaled by a correction coefficient of 1.25 to approach the converged values.

\begin{figure}[!t]
\center{\includegraphics
{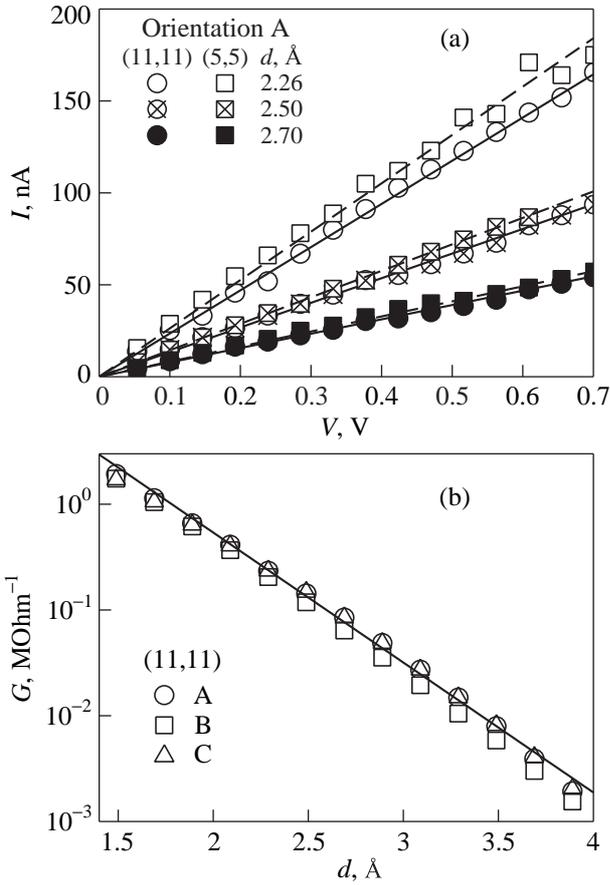}}
\caption{(a) Calculated tunneling current $I$ between the ends of the coaxial (11,11) nanotubes and (5,5) nanotubes as a function of voltage $V$ for relative orientation A of the nanotubes at the distances $d$ between the nanotube ends of 2.26, 2.50 and 2.70~\AA{}. The lines show the linear approximations $I = GV$ of these dependences, where $G$~[MOhm$^{-1}$] for different $d$ is: 0.235 (2.26~\AA{}), 0.134 (2.50~\AA{}), and 0.078 (2.70~\AA{}) for the (11,11) nanotubes; 0.263 (2.26~\AA{}), 0.144 (2.50~\AA{}), and 0.082 (2.70~\AA{}) for the (5,5) nanotubes. (b) Calculated tunneling conductance $G$ between the ends of the coaxial (11,11) nanotubes as a function of the distance $d$ between the nanotube ends along their common axis for different orientations (A, B, C) of the nanotubes. The line shows the exponential approximation of the dependence for orientation A of the nanotubes: $G\,[\mathrm{MOhm}^{-1}] = (155\pm21)\exp[-(2.83\pm0.05)d\,[\mathrm{\AA}]]$.} \label{figures:Fig04CalculatedTunnelingCurrent11}
\end{figure}

The calculations of tunneling current between the (11,11) nanotubes with relative orientation A and at the distances between their ends of 2.26~\AA{}, 2.50~\AA{} and 2.70~\AA{} demonstrate that the $I$--$V$ curves of the nanotube contact are linear at voltages below 0.7~V (Fig.~\ref{figures:Fig04CalculatedTunnelingCurrent11}a). At these voltages, the tunneling conductance is found as the current to voltage ratio. To study the dependence of tunneling conductance at different distances between the nanotubes the calculations of the tunneling current have been performed for the voltage of 0.1~V. The obtained curves (Fig.~\ref{figures:Fig04CalculatedTunnelingCurrent11}b) follow the semiclassical rule $G(d) = G_0\exp(-\alpha d)$ \cite{Landau89}, where $G_0 = (155\pm21)$~MOhm$^{-1}$ and $\alpha = (2.83\pm0.05)$~\AA$^{-1}$ for relative orientation A of the nanotubes. The differences between the tunneling conductances for relative orientations B-A and C-A are within 10\% and 30\%, respectively.

For comparison, the $I$--$V$ curves calculated for the (5,5) nanotubes with relative orientation A and at the distances between their ends of 2.26~\AA{}, 2.50~\AA{} and 2.70~\AA{} are also shown in Fig.~\ref{figures:Fig04CalculatedTunnelingCurrent11}a. It is seen that the tunneling conductance between the ends of the (5,5) nanotubes is almost the same as for the (11,11) nanotubes. 
Though for small-diameter armchair carbon nanotubes the work function is known to decrease with decreasing the nanotube diameter, this dependence is rather weak and the difference in the work functions for the (5,5) and (11,11) nanotubes is very small \cite{ZhaoHanLu02,ShanCho05,SuLeungChan07}. The difference in the circumference lengths of the nanotubes does not affect the tunneling current either. As follows from the Landauer formula \cite{Fonseca03, Anantram06}, the dependence of conductance of metallic systems on the cross-section size is related to changes in the number of subbands crossing the Fermi level. However, for all metallic single-walled nanotubes, this number is equal to 2  irrespective of their diameter \cite{Anantram06}. This explains why no significant difference is observed in the tunneling currents for the (5,5) and (11,11) nanotubes and supports adequacy of the calculations with the converged parameters.

\section{Characteristics of the system as a force sensor}
\label{ForceSensorSection}

\begin{figure}[h]
\center{\includegraphics
{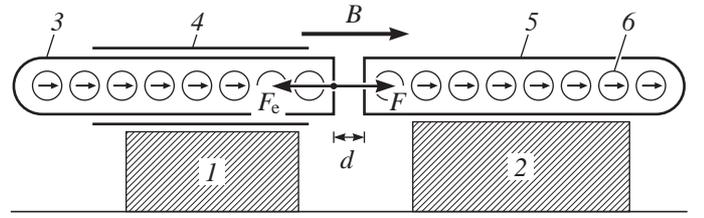}}
\caption{A general scheme of the force sensor based on interaction between carbon ends. The electrodes (\emph{1}) and (\emph{2}), movable inner wall (\emph{3}) and fixed outer wall (\emph{4}) of the first double-walled nanotube, and second fixed single-walled nanotube (\emph{5}) are indicated. The arrows correspond to a positive value of the measured external force $F$ and the force of van der Waals interaction between nanotubes ends $F_{\mathrm{e}}$. The nanotubes filled with magnetic endofullerenes (\emph{6}) can be used for measurements of the magnetic field $B$. Endofullerenes are fixed inside the nanotubes.} \label{figures:Fig05ForceSensorGeneral}
\end{figure}

The general scheme of the force sensor based on measurements of conductance between ends of two coaxial carbon nanotubes is shown in Fig.~\ref{figures:Fig05ForceSensorGeneral}. The first nanotube is a double-walled nanotube (DWNT) with a movable inner wall (\emph{3}) and a fixed outer wall (\emph{4}) positioned on the first electrode (\emph{1}) and the second one is a fixed single-walled nanotube (\emph{5}) positioned on the second electrode (\emph{2}). The operation of this force sensor is determined by the balance of forces that act on the movable wall (\emph{3}): the force of van der Waals interaction between the nanotubes ends, $F_\mathrm{e}$, the force of van der Waals interaction between the movable wall (\emph{3}) and the fixed wall (\emph{4}), $F_\mathrm{w}$, and the measured external force $F$. 
We consider here the operation of the force sensor in the case where the measured external force $F$ is directed along the common axis of the nanotubes. When the measured external force $F$ is applied to the movable wall (\emph{3}), the distance between the ends of the oppositely located nanotubes changes. The change of this distance causes the change of the tunneling conductance between the nanotube ends and, therefore, the change of the conductance between the electrodes (\emph{1}) and (\emph{2}). Thus, the external force $F$ can be determined through the measurements of the conductance between the electrodes (\emph{1}) and (\emph{2}). Evidently the energy of van der Waals interaction between the walls (\emph{3}) and (\emph{4}) is proportional to the overlap length of the walls. Therefore, if the inner wall is telescopically extended only from one end of the outer wall, the force $F_\mathrm{w}$ retracts the inner wall back into the outer wall. Calculations show that in this case the force $F_\mathrm{w}$ exceeds the maximal value of the force $F_\mathrm{e}$ of interaction between the nanotubes ends \cite{PopovBichoutskaiaLozovikKulish07}. In another case where the length of the inner wall exceeds the length of the outer wall in such a way that both ends of the inner wall are telescopically extended from the outer wall, the overlap length of the walls is constant and the value of force $F_\mathrm{w}$ is determined only by the relative position of the inner and outer walls and as discussed below can be significantly less than the force $F_\mathrm{e}$. It is the case that is proposed here for application in the general force sensor scheme (see Fig.~\ref{figures:Fig05ForceSensorGeneral}).

Let us first discuss the influence of the force $F_\mathrm{w}$ of interaction between the walls of the first
nanotube on the operational characteristics of the sensor. The theoretical analysis \cite{Barreiro08,
Saito01, LozovikMinoginPopovPhysLett03, LozovikMinoginPopovJETPLett03, Kolmogorov00, Damnjanovic02,
Vukovic03, Damnjanovic03, Belikov04, BichoutskaiaPopov05, Charlier93, KwonTomanek98, Palser99,
BichoutskaiaHeggie06, PopovLozovikSobennikovKnizhnik09, KolmogorovCrespi05, Carlson07, RomanPerez09, Hod10,
Hod13, PopovLebedevaKnizhnik13} has shown that there are three types of pairs of adjacent nanotube walls with
basically different characteristics of the force $F_\mathrm{w}$ of interaction between the walls:
(i)~commensurate walls at least one of which is chiral, (ii)~commensurate nonchiral walls (i.e. both armchair
or both zigzag walls), and (iii)~incommensurate walls. For infinite commensurate walls at least one of which
is chiral, corrugations in the dependence of interwall interaction energy on the relative position of the
walls are smaller than the accuracy of calculations \cite{Vukovic03, Damnjanovic03, Belikov04, RomanPerez09}.
This is so because incompatibility of helical symmetries of the walls provides that only very high Fourier
harmonics of interaction energy between an atom of one of the walls and the second wall contribute to
corrugations of the potential energy relief, whereas the contributions of other harmonics corresponding to
different atoms are completely compensated \cite{Damnjanovic02}. Calculations in the framework of the density
functional theory (DFT) show that for finite commensurate chiral walls corrugations in the dependence of
interwall interaction energy on the relative position of the walls are determined by contributions of edges
and the magnitude of these corrugations is 3--50~meV, which is one-two orders of magnitude less than the
absolute value of interaction energy between the nanotubes ends \cite{PopovLebedevaKnizhnik13}
(Fig.~\ref{figures:Fig02CalculatedChangesE}b). For incommensurate walls, the magnitude of the corrugations
fluctuates near some average value and does not increase with increasing the overlap length of the walls
\cite{Kolmogorov00}. In this case, DFT calculations also give the corrugations of the potential energy relief
within 50~meV \cite{PopovLebedevaKnizhnik13}. Therefore, in both cases discussed above the force
$F_\mathrm{w}$ of interaction between the walls is considerably less than the force $F_\mathrm{e}$ of
interaction between the nanotubes ends. The considerable barriers to relative sliding of the walls along the
nanotube axis are found only for the case of commensurate nonchiral walls using DFT calculations
\cite{BichoutskaiaPopov05, Charlier93, KwonTomanek98, Palser99, BichoutskaiaHeggie06,
PopovLozovikSobennikovKnizhnik09, RomanPerez09, Hod10, Hod13, PopovLebedevaKnizhnik13} and calculations based
on tight-binding \cite{Carlson07} and empirical \cite{KangLeeKimChoi09, Kolmogorov00, Damnjanovic02,
Vukovic03, Damnjanovic03, Belikov04, KolmogorovCrespi05} potentials. For DWNTs with arbitrary chirality
indexes of walls, the probability that the walls are commensurate and nonchiral is very low. For example,
such DWNTs have not been found among 141 DWNTs with chirality indexes determined by electron diffraction
\cite{Hirahara06}. Thus we consider the case where the operation of the force sensor is determined by the
balance $F_\mathrm{e}=F$ of only two oppositely directed forces, the force $F_\mathrm{e}$ and the measured
external force $F$.

\begin{figure}[!t]
\center{\includegraphics
{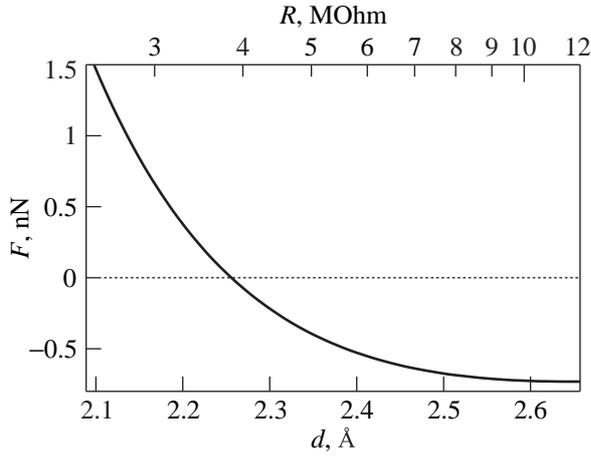}}
\caption{Measured external force $F$ as a function of the distance $d$ and of the resistance $R = 1/G$ for the force sensor based on the coaxial (11,11) nanotubes.} \label{figures:Fig06ExternalForceDistance}
\end{figure}

The calculated dependence of the measured external force $F$ on the distance between the (11,11) nanotubes ends is shown in Fig.~\ref{figures:Fig06ExternalForceDistance}. For the external force $F$ which is directed so that it leads to an increase in the distance between the nanotubes ends, the maximal value that leads to a finite displacement of the movable wall is 0.73~nN and corresponds to the distance between the nanotubes ends $d = 2.65$~\AA. For greater values of the external force $F$, the stable position of the movable wall is not possible. Therefore, the range of forces which can be measured by the considered force sensor is within 0.5~nN. The calculated dependences of the measured external force $F$ on the contact resistance $R = 1/G$ between the nanotubes ends for this range of forces is shown in Fig.~\ref{figures:Fig06ExternalForceDistance}. The sensitivity of the force sensor for small values of the measured force $F$ is $\Delta R/\Delta F = 1.9$~MOhm$/$nN.

Let us also discuss the possible nature of the measured force $F$. The accelerometer based on the dependence of characteristics of telescopic oscillations of the DWNT inner wall on the inertial force acting on it was considered in Ref.~\cite{WangJiang08}. It was proposed to improve the accelerometer sensitivity by increasing the inner wall mass as a result of encapsulation of a metal core. The scheme of the force sensor suggested here allows to increase the mass of the movable part of the accelerometer and thus its sensitivity by attaching nanoobjects to the outer surface of the inner wall. Another application of the force sensor can be also realized if a molecule or nanoobject is adsorbed on the movable inner wall in the region where it does not overlap with the fixed outer wall. Namely, measurements of the force acting on the molecule or nanoobject in the presence of an electric field would allow to determine its polarizability or electric dipole moment. The possibility to measure a magnetic force between ends of coaxial nanotubes filled with magnetic endofullerenes is considered in the following Section.

\section{Magnetic field measurements}
\label{MagneticForceSection}

\subsection{Zero temperature}
Let us consider the possibility to measure a magnetic force between two coaxial nanotubes filled with (Ho$_3$N)@C$_{80}$ magnetic endofullerenes with magnetic moments $M = 21\mu_\mathrm{B}$ with the use of the proposed force sensor (see Fig.~\ref{figures:Fig05ForceSensorGeneral}); $\mu_\mathrm{B} = 57.88$~\textmu eV$/$T is the Bohr magneton. We imply that at any temperature endofullerenes are fixed inside the nanotubes (due to the van der Waals forces). When the magnetic moments of the endofullerenes are codirectional, the attractive magnetic force $F_{{\mathrm{m}}(ij)}$ between two endofullerenes in the left and right nanotubes is maximal in the case where their magnetic moments are directed along the nanotube axis and is given by \cite{White07}
\begin{equation}\label{eq104}
   F_{\mathrm{m}(ij)} = 6\mu_0\frac{M^2}{r_{ij}^4},
\end{equation}
where $r_{ij}$ is the distance between the endofullerenes, $\mu_0 = 1.257$~\textmu H$/$m is the magnetic constant, $i$ and $j$ are the numbers of the endofullerenes in the left and right nanotubes, respectively, starting from the nearest nanotube ends.

\begin{figure}[!t]
\center{\includegraphics
{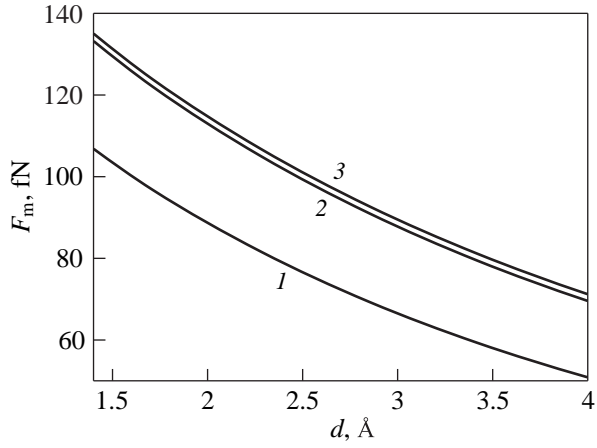}}
\caption{Calculated magnetic force between the coaxial (11,11) nanotubes filled with the (Ho$_3$N)@C$_{80}$ magnetic endofullerenes as a function of the distance $d$ between the nanotube ends along the nanotube axis for the cases of one (line \emph{1}), five (line \emph{2}) and 50 endofullerenes (line \emph{3}).} \label{figures:Fig07MagneticForceDistance}
\end{figure}

The total force $F_\mathrm{m}$ of magnetic attraction between two coaxial nanotubes filled with endofullerenes with the magnetic moments directed parallel to the nanotube axis takes the form
\begin{equation}\label{eq105}
   F_\mathrm{m} = 6\mu_0M^2\sum_{i=0}^{N_\mathrm{l}-1}\sum_{j=0}^{N_\mathrm{r}-1}\frac{1}{(d + d_\mathrm{ef} + id_\mathrm{ef} + jd_\mathrm{ef})^4}.
\end{equation}
Here $d + d_\mathrm{ef}$ is the distance between centers of the first endofullerenes in the nanotubes, $N_\mathrm{l}$ and $N_\mathrm{r}$ are the total numbers of endofullerenes in the left and right nanotubes, $d_\mathrm{ef} \approx (8 + 3.4)$~\AA{} is the distance between centers of the nearest endofullerenes in the nanotubes, where 8~\AA{} corresponds to the diameter of the (Ho$_3$N)@C$_{80}$ endofullerene, 3.4~\AA{} is the distance between the surfaces of adjacent endofullerenes which we assume to be equal to the interlayer distance of graphite. The calculated magnetic force between the (11,11) nanotubes filled with the (Ho$_3$N)@C$_{80}$ magnetic endofullerenes as a function of the distance $d$ between the nanotubes ends is presented in Fig.~\ref{figures:Fig07MagneticForceDistance}.

Figure~\ref{figures:Fig07MagneticForceDistance} shows that the maximal value of magnetic force is three orders of magnitude smaller than the range of forces which can be measured by the force sensor. Therefore, small forces should be measured to determine magnetic fields. However, starting from 1972, the methods of resistance measurements with the accuracy less than 0.01~ppm have been elaborated \cite{Harvey72}. We show below that this accuracy is sufficient for measurements of magnetic fields and, moreover, these measurements can be done with a high spacial resolution. Figure~\ref{figures:Fig07MagneticForceDistance} demonstrates that at the distances between nanotubes comparable to the endofullerene diameter only several magnetic endofullerenes near the interacting ends of the coaxial nanotubes are required to achieve the magnetic force close to its maximal possible value for the infinite nanotubes completely filled with the endofullerenes. Therefore, the length of the force sensor for magnetic field measurements is restricted only by the possibilities of modern nanotechnology.

Note also that operational principles of proposed force sensor are based on the assumption that the energy of interaction between the coaxial nanotubes has a minimum at the distance of several angstroms between their ends (equilibrium position of movable wall at the distance $d_0$). Thus, the presence of the measured force acting on the movable wall leads to displacement of the equilibrium position that can be detected by measuring of the tunneling conductance between the nanotube ends. In the analysis above, we take into account only the van der Waals interaction between the coaxial nanotubes. However, the presence of other forces of interaction between the coaxial nanotubes filled by endofullerenes does not make the operation of the force sensor impossible if the distance $d_0$ does not change considerably. For example, the possibility of charge transfer from the (Dy$_3$N)@C$_{80}$ \cite{Shiozawa06} and Dy@C$_{82}$ \cite{KitauraOkimoto07} endofullerenes inside the nanotube to the nanotube wall was proposed, which can lead to the Coulomb repulsion between the endofullerenes. Nevertheless, this Coulomb repulsion between the endofullerenes inside the same nanotube does not prevent filling of the nanotube by the endofullerenes. Therefore this repulsion does not exceed the van der Waals attraction between the endofullerene and the nanotube. For this reason, we believe that the Coulomb repulsion between the endofullerenes inside the coaxial nanotubes should not exceed the van der Waals attraction between their ends either and thus cannot make the operation of the magnetic field sensor impossible.

\subsection{Finite temperature}
Let us consider the magnetic interaction between coaxial nanotubes filled with magnetic endofullerenes in the case where a magnetic field $B$ is applied in the direction parallel to the nanotube axis at finite temperature $T$. We assume that the magnetic endofullerenes inside the nanotubes have a paramagnetic state in the absence of the external magnetic field. In this case, statistical averaging of the attractive magnetic forces \cite{White07} between the endofullerenes over all orientation directions of their magnetic moments yields
\begin{equation}\label{eq108}
   F_\mathrm{m} = 6\mu_0M^2[L(\beta)]^2 \sum_{i=0}^{N_\mathrm{l}-1}\sum_{j=0}^{N_\mathrm{r}-1} \frac{1}{(d + d_\mathrm{ef} + id_\mathrm{ef} + jd_\mathrm{ef})^4},
\end{equation}
where $L(\beta) = \coth(\beta) - \beta^{-1}$ is the Langevin function and $\beta = BM/k_\mathrm{B}T$. (Note that at $B\to0$ the force $F_\mathrm{m}\to0$, because without external magnetic field the magnetic moments of endofullerenes are randomly oriented.)

As seen from this formula, the range of magnetic fields that can be measured using the sensor is sensitive to temperature. This is so because the magnetic moments of the endofullerenes deviate from the direction of the magnetic field due to thermal fluctuations. At liquid helium temperature the range of measured magnetic fields is within 10~T. At room temperature the maximum measured magnetic field strength can achieve hundreds of tesla. However, it should be mentioned that at ultrahigh magnetic fields, the expression (\ref{eq108}) is invalid since it does not take into account nonlinear and quantum effects, such as the dependence of the magnetic moments on the magnetic field \cite{MoonKimOh00} and quantization of electronic states \cite{Lisitsa87}.

\begin{figure}[!t]
\center{\includegraphics
{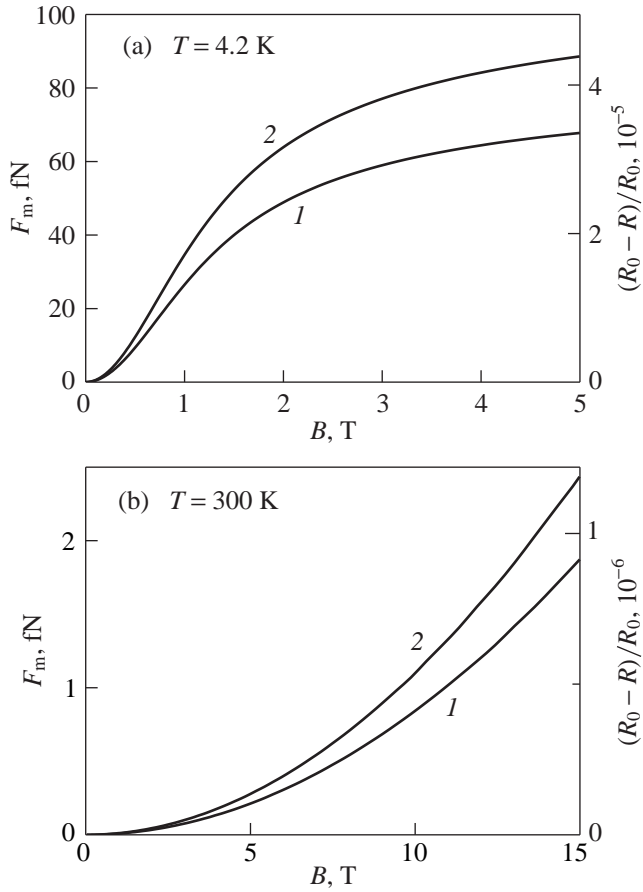}}
\caption{Calculated force $F_\mathrm{m}$ of magnetic attraction between the coaxial (11,11) nanotubes filled with the (Ho$_3$N)@C$_{80}$ magnetic endofullerenes and relative change of the contact resistance $(R_0 - R)/R_0$ of the nanotubes as functions of the magnetic field strength $B$. The magnetic field $B$ is parallel to the nanotube axis. The nanotubes contain 1 (line \emph{1}) and 50 (line \emph{2}) endofullerenes. The distance between the nanotubes ends is $d = 2.26$~\AA. Temperature is 4.2~K (a) and 300~K (b).}\label{figures:Fig08MagneticForceMagneticField}
\end{figure}

The dependences of the magnetic force $F_\mathrm{m}$ on the applied magnetic field $B$ at liquid helium temperature $T = 4.2$~K and room temperature $T = 300$~K for $B < 15$~T (where the nonlinear and quantum effects are not essential and expression (\ref{eq108}) is adequate) are presented in Fig.~\ref{figures:Fig08MagneticForceMagneticField}. Correlating this dependence with the calculated dependences of the tunneling conductance and interaction force between the nanotubes ends on the distance between them, we obtain the relative change $(R_0 - R)/R_0$ ($R_0 \approx 3.9$~MOhm is the contact resistance at $B = 0$) of the contact resistance of the nanotubes as a function of the applied magnetic field $B$ (see Fig.~\ref{figures:Fig08MagneticForceMagneticField}). The calculated relative change of the contact resistance at liquid helium temperature for a magnetic field of several tesla is a few orders of magnitude greater than the accuracy of resistance measurements, which is less than 0.01~ppm \cite{Harvey72}. Thus, temperatures $T < 300$~K and magnetic fields $B < 15$~T are suitable conditions for operation of the proposed magnetic field sensor with the accuracy of measurements about 1\%.

Because of the small size of the sensor, it not only holds promise for measurements of local magnetic fields with the high spatial resolution but also a small response time of the sensor can be expected. To estimate the response time of the device we consider the contact between the nanotubes as a parallel $RC$-circuit with the resistance $R \approx 3.9$~MOhm and the capacitance $C = \varepsilon_0S/d$, where $d = 2.26$~\AA{} is the distance between the nanotubes ends, $\varepsilon_0 = 8.854\cdot10^{-12}$~F/m and $S$ is the contact area. For the (11,11) nanotubes $S = (33a_\mathrm{CC})^2/(4\pi) \approx 175$~\AA$^2$, $a_\mathrm{CC} = 1.42$~\AA{} is the bond length of graphene. The response time will be $\tau_\mathrm{r}= RC\approx 0.3~\mathrm{ ps}$. With the range of measured magnetic fields of hundreds of tesla at room temperature, this makes the proposed sensor perspective for applications in experiments related to impulse ultrahigh magnetic fields.

\section{Conclusions}
\label{ConclusionsSection}

We have performed atomistic calculations of the tunneling current through the ends of coaxial (11,11) and (5,5) carbon nanotubes terminated by hydrogen atoms as a function of the distance between these ends and the voltage applied. The Green function formalism for systems with semi-infinite periodic electrodes has been applied. The calculated $I$--$V$ curves are linear at voltages below 0.7~V and distances between the nanotube ends in the range from 2.25~\AA{} to 2.7~\AA. At these voltages, the tunneling conductance, $G$, is obtained as the current to voltage ratio. It is found that the dependence of the conductance $G$ on the distance $d$ between the nanotube ends follows the semiclassical rule $G(d) = G_0\exp(-\alpha d)$ with $G_0 = (155\pm21)$~MOhm$^{-1}$ and $\alpha = (2.83\pm0.05)$~\AA$^{-1}$ for the (11,11) nanotubes. The differences in the tunneling conductances for different relative orientations of the nanotubes are within 30\%. It is also revealed that the tunneling conductance between the ends of the (5,5) nanotubes is only slightly higher than for the (11,11) nanotubes, i.e. the tunneling current weakly depends on the nanotube diameter.

The interaction between the same (11,11) carbon nanotubes is studied using semiempirical potentials as a function of the distance between the nanotube ends and the angle of their relative orientation. It is shown that the optimal orientation of the nanotubes corresponds to the case when hydrogen atoms of one of the nanotubes are located in the middle between hydrogen atoms of the neighboring edge of the second nanotube. The equilibrium distance between the nanotubes with this relative orientation is found to be 2.26~\AA. The barrier for relative rotation of the coaxial nanotubes is found to be about 0.07~eV. Thus, this rotation is completely frozen at liquid helium temperature, whereas at room temperature it becomes possible.

The calculated dependences of the interaction energy and tunneling conductance on the distance between the nanotube ends are used to obtain the contact resistance between the coaxial nanotubes as a function of the force of interaction between them. The general scheme and operational principles of the force sensor based on measurements of the contact resistance are proposed. This force sensor can be used as an accelerometer. A molecule or nanoobject can be adsorbed on the movable inner wall in the region where it does not overlap with the fixed outer wall. Measurements of the force acting on the molecule or nanoobject in the presence of an electric field would allow to determine its polarizability or electric dipole moment.

Particularly, in the present paper we have considered the possibility to detect magnetic field trough measurements of the magnetic force between the ends of coaxial nanotubes by the example of the (11,11) nanotubes filled with the (Ho$_3$N)@C$_{80}$ endofullerenes having the largest observed magnetic moment. In this case, the range of magnetic fields that can be measured using the sensor corresponds to several tesla at liquid helium temperature and several hundreds of tesla at room temperature. Since the sensor has small dimensions its response time is estimated to be less than picosecond and, thus, it can be used for measurements of impulse ultrahigh magnetic fields both with high spatial and temporal resolution. The proposed force sensor can be also applied for measurements of magnetic moments of endofullerenes or other magnetic molecules inside single-walled carbon nanotubes.

\section*{Acknowledgments}

A.M.P. and Y.E.L. acknowledges Russian Foundation of Basic Research (12-02-900241-Bel) and Global Research Outreach Program. I.V.L. acknowledges support by the Marie Curie International Incoming Fellowship within the 7th European Community Framework Programme (Grant Agreement PIIF-GA-2012-326435 RespSpatDisp), Grupos Consolidados del Gobierno Vasco (IT-578-13) and the computational time on the Supercomputing Center of Lomonosov Moscow State University and the Multipurpose Computing Complex NRC ``Kurchatov Institute''. N.A.P., A.I.S., S.A.V. and S.V.R. acknowledge support by the Belarusian Republican Foundation for Fundamental Research (Grant No.~F14R-088) and Belarusian scientific program ``Convergence''.







\end{document}